\documentclass[prl,aps,twocolumn,superscriptaddress]{revtex4-1}

\usepackage{amsmath,amssymb,amsfonts,bbm,color,graphicx,tabularx}

\usepackage[unicode=true,colorlinks=true]{hyperref}
\usepackage{chemformula}
\hypersetup{linkcolor=blue,citecolor=blue,urlcolor=blue}

\begin{document}

\title{Microscopic models for Kitaev's sixteenfold way of anyon theories}

\author{Sreejith Chulliparambil}
\affiliation{Institut f\"ur Theoretische Physik and W\"urzburg-Dresden Cluster of Excellence ct.qmat, Technische Universit\"at Dresden, 01062 Dresden, Germany}
\affiliation{Max-Planck-Institut f{\"u}r Physik komplexer Systeme, N{\"o}thnitzer Stra{\ss}e 38, 01187 Dresden, Germany}

\author{Urban F. P. Seifert}
\affiliation{Institut f\"ur Theoretische Physik and W\"urzburg-Dresden Cluster of Excellence ct.qmat, Technische Universit\"at Dresden, 01062 Dresden, Germany}

\author{Matthias Vojta}
\affiliation{Institut f\"ur Theoretische Physik and W\"urzburg-Dresden Cluster of Excellence ct.qmat, Technische Universit\"at Dresden, 01062 Dresden, Germany}

\author{Lukas Janssen}
\affiliation{Institut f\"ur Theoretische Physik and W\"urzburg-Dresden Cluster of Excellence ct.qmat, Technische Universit\"at Dresden, 01062 Dresden, Germany}

\author{Hong-Hao Tu}
\email{hong-hao.tu@tu-dresden.de}
\affiliation{Institut f\"ur Theoretische Physik and W\"urzburg-Dresden Cluster of Excellence ct.qmat, Technische Universit\"at Dresden, 01062 Dresden, Germany}

\begin{abstract}
In two dimensions, the topological order described by $\mathbb{Z}_2$ gauge theory coupled to free or weakly interacting fermions with a nonzero spectral Chern number $\nu$ is classified by $\nu \; \mathrm{mod}\; 16$ as predicted by Kitaev [\href{https://doi.org/10.1016/j.aop.2005.10.005}{Ann.\ Phys.\ \textbf{321}, 2 (2006)}]. Here, we provide a systematic and complete construction of microscopic models realizing this so-called sixteenfold way of anyon theories. These models are defined by $\Gamma$ matrices satisfying the Clifford algebra, enjoy a global $\mathrm{SO}(\nu)$ symmetry, and live on either square or honeycomb lattices depending on the parity of $\nu$. We show that all these models are exactly solvable by using a Majorana representation and characterize the topological order by calculating the topological spin of an anyonic quasiparticle and the ground-state degeneracy. The possible relevance of the $\nu=2$ and $\nu=3$ models to materials with Kugel-Khomskii-type spin-orbital interactions is discussed.
\end{abstract}

\date{\today}

\maketitle

In recent years, topological phases of matter have attracted tremendous interest partially due to their potential applications in fault-tolerant quantum information processing~\cite{nayak2008,kitaev2003}. For building quantum hardware with topological protection, it is crucial to find suitable microscopic models stabilizing the topological phases, and then identify and synthesize suitable materials, or build appropriate quantum simulators.

The celebrated example where such a program has been carried out is Kitaev's honeycomb model~\cite{kitaev2006b}. This model is a rare example for which non-Abelian topological order is known to exist through an exact solution. The model's highly anisotropic magnetic interactions indeed emerge in certain $4d$ and $5d$ transition-metal compounds with strong spin-orbit interactions~\cite{jackeli2009}. These compounds, nowadays termed Kitaev materials~\cite{trebst2017,janssen2019}, also have other interactions apart from the Kitaev interaction. It is thus not obvious at all whether the non-Abelian topological order survives in such complicated situations. A prominent candidate is $\alpha$-RuCl$_3$, for which the observed approximately half-quantized thermal Hall conductance hints at Majorana edge states and hence non-Abelian (Ising) topological order in the bulk~\cite{kasahara2018,yokoi2020}.

It is very natural to look further for other topological phases beyond the Ising topological order. In Kitaev's seminal work~\cite{kitaev2006b}, it was proposed that two-dimensional (2D) topological superconductors (by which we refer to free or weakly interacting fermions with broken number conservation) with a Chern number $\nu$ coupled to $\mathbb{Z}_2$ gauge fields give rise to a series of topological orders classified by $\nu \; \mathrm{mod}\; 16$, which was termed the ``sixteenfold way.'' The Abelian and non-Abelian topological orders realized in Kitaev's honeycomb model correspond to the first two instances ($\nu=0$ and $1$). There have been multiple attempts~\cite{kells2011,zhang2020,fuchs2020} to generalize Kitaev's honeycomb model, with the aim of finding microscopic models realizing all 16 anyon theories. However, a complete solution has not been achieved so far, mostly due to the difficulty of engineering topological superconductors with large Chern numbers. An alternative approach adopted in Ref.~\cite{tu2013a} conjectured that the conformal field theory (CFT) for describing the gapless edge excitations in Kitaev's sixteenfold way is the SO($\nu$)$_1$ Wess-Zumino-Witten model and used the bulk-edge correspondence~\cite{moore1991,nielsen2012} to construct bulk wave functions and their parent Hamiltonians. However, the drawback of this approach is that the parent Hamiltonians have long-range interactions and only the ground state is known, which prohibits access to physical properties from this microscopic construction.

In this work, we provide a systematic and complete construction of microscopic models realizing Kitaev's sixteenfold way. For Abelian (even $\nu$) and non-Abelian (odd $\nu$) theories, the models are defined on the square and honeycomb lattices, respectively, and have short-range interactions defined with $\Gamma$ matrices satisfying the Clifford algebra. All these models are exactly solvable in terms of a Majorana fermion representation, where the model describes $\nu$ species of itinerant Majorana fermions, each of which has a Chern number $1$, coupled to a static $\mathbb{Z}_2$ gauge field.

The reduction of an interacting spin Hamiltonian to free fermions coupled to a static gauge field, and the selection rules following from the conservation of the gauge field, admit rare reliable insights into the key characteristics of a quantum spin liquid, as has been demonstrated for Kitaev's honeycomb model: Static~\cite{baskaran2007} and dynamic~\cite{knolle2014} spin correlation functions can be evaluated exactly, exhibiting key signatures of fractionalization, and finite-temperature properties can be obtained using unbiased Quantum Monte Carlo methods~\cite{motome2015}.
Further, the model's exact solution allows one to reveal hidden string orders~\cite{feng2007} and evaluate the topological entanglement entropy $S_{\mathrm{top}}=\ln 2$~\cite{yao2010}. Since the exact solutions of the microscopic models put forward in this work share the same features as the Kitaev's honeycomb model, the methods to obtain these results directly carry over.
%For example, dynamical spin correlations for the $\nu = 3$ model have recently been obtained~\cite{natori2020}.

Based on the exact solution of the microscopic model, we characterize the topological order by showing that the topological spin of an anyonic quasiparticle is $\theta = \nu\pi/8$ and the ground-state degeneracy is four (three) for Abelian (non-Abelian) models on the torus. Finally, we discuss the relevance of the $\nu=2$ and $\nu=3$ models to Kugel-Khomskii type spin-orbital systems. These models support chiral spin liquids which have the same topological order as the Laughlin state at $1/4$ filling and the Moore-Read state at unit filling in fractional quantum Hall systems, respectively.

\begin{figure}
\includegraphics[width=\linewidth]{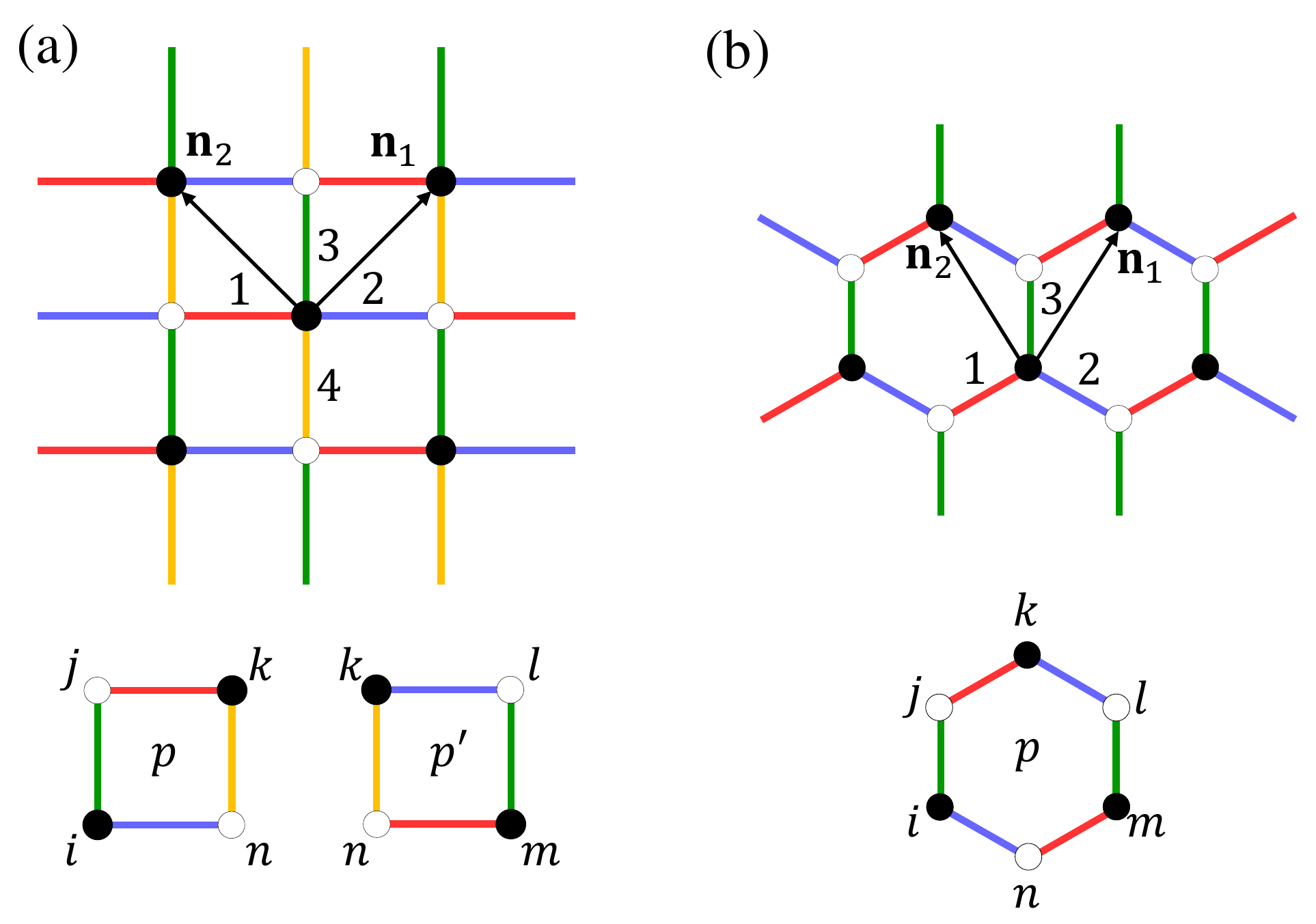}
\caption{Schematics of (a) square and (b) honeycomb lattices. The black (white) dots correspond to the A (B) sublattice. The four (three) types of links on the square (honeycomb) lattice are denoted by different color codes. The square and hexagon illustrate the definition of the plaquette operators. The primitive vectors for the square (honeycomb) lattice are $\mathbf{n}_{1,2} = (\pm \frac{1}{\sqrt{2}},\frac{1}{\sqrt{2}})$ [$\mathbf{n}_{1,2} = (\pm \frac{1}{2},\frac{\sqrt{3}}{2})$].
}
\label{fig:lattice}
\end{figure}

{\em Models.}---For each $\nu$, we construct an exactly solvable lattice model. This family of models is defined on the square (honeycomb) lattice for even (odd) $\nu$, which, respectively, starts with Wen's plaquette model~\cite{wen2003a,wen2003b} and Kitaev's honeycomb model~\cite{kitaev2006b} (abbreviated as $\nu = 0$ and $\nu = 1$ models hereafter). For $\nu = 2q$ and $2q+1$ ($q \in \mathbb{N}_0$), the two models have a $2^{q+1}$-dimensional Hilbert space at each lattice site. Accordingly, the local operators for constructing the Hamiltonian are given by the generators of a $2^{q+1}$-dimensional representation of the Clifford algebra, $\Gamma^\alpha$ ($\alpha=1,\ldots,2q+3$), which are Hermitian and satisfy $\{\Gamma^\alpha,\Gamma^\beta\} = 2\delta_{\alpha\beta}$, as well as their commutators $\Gamma^{\alpha \beta} = \frac{i}{2}[\Gamma^\alpha,\Gamma^\beta]$. While the $\nu = 0$ model requires a separate definition, the Hamiltonian for $\nu>0$ can be written as
\begin{equation}
    H  = - \sum_{\langle ij\rangle_{\gamma}} J_{\gamma}
    \left(\Gamma_{i}^{\gamma}\Gamma_{j}^{\gamma} + \sum_{\beta=\gamma_{\textrm{m}}+1}^{2q+3} \Gamma_{i}^{\gamma\beta} \Gamma_{j}^{\gamma\beta}\right),
\label{eq:Hamiltonian}
\end{equation}
where $\langle ij\rangle_{\gamma}$ denotes for different types of links between neighboring sites (see Fig.~\ref{fig:lattice}) and $\gamma_{\textrm{m}} = 4$ (3) for the square (honeycomb) lattice. Among this family of models, the $\nu = 2$ and $\nu = 3$ cases have been considered in Ref.~\cite{nakai2012} and Refs.~\cite{yao2011,natori2020}, respectively.

For $\nu>1$, the Hamiltonian (\ref{eq:Hamiltonian}) commutes with $\sum_j\Gamma_{j}^{\alpha}$ and $\sum_j\Gamma_{j}^{\alpha\beta}$, where $\alpha$ and $\beta$ range from 5 (4) to $2q+3$ for the square (honeycomb) model. As the latter operators form a closed SO($\nu$) algebra, the Hamiltonian (\ref{eq:Hamiltonian}) enjoys a global SO($\nu$) symmetry~\cite{appendix}. Furthermore, the Hamiltonian (\ref{eq:Hamiltonian}) has plaquette operators as its local integrals of motion, generalizing the situation for the $\nu = 1$ model~\cite{kitaev2006b}. The plaquette operator for the honeycomb lattice is defined on each hexagon as $W_{p}=\Gamma^{23}_i \Gamma^{31}_j \Gamma^{12}_k \Gamma^{23}_l \Gamma^{31}_m \Gamma^{12}_n$ [see Fig.~\ref{fig:lattice}(b)], whereas the square lattice has two types of squares, as shown in Fig.~\ref{fig:lattice}(a), with plaquette operators defined as $W_p = -\Gamma^{23}_i \Gamma^{31}_j \Gamma^{14}_k \Gamma^{42}_n$ and $W_{p'} = -\Gamma^{42}_k \Gamma^{23}_l \Gamma^{31}_m \Gamma^{14}_n$~\footnote{The $\nu=0$ model is restored by defining $H = -J\sum_{p} W_p -J\sum_{p'} W_{p'}$ with $\Gamma^{23}=\Gamma^{14}=\sigma^x$ and $\Gamma^{31}=\Gamma^{42}=\sigma^y$.}, respectively.
These plaquette operators are special cases of the Wilson loop operators, which are written as~\cite{wu2009}
\begin{equation}
    W_{\mathcal{C}} = (-1)^{N/2-1} \Gamma^{\gamma_N\gamma_1}_{i_1} \Gamma^{\gamma_1 \gamma_2}_{i_2} \cdots \Gamma^{\gamma_{N-1}\gamma_N}_{i_N}.
\label{eq:loop}
\end{equation}
Here, $\mathcal{C}$ stands for a non-intersecting loop with $N$ successive neighboring sites (clockwisely labeled by $i_1,\ldots,i_N$), and the superscripts of the $\Gamma$ operators correspond to two types of links on the loop crossing one site. All loop operators $W_{\mathcal{C}}$ commute with the Hamiltonian (\ref{eq:Hamiltonian}) and have eigenvalues $\pm 1$. When the system is defined on a topologically nontrivial manifold (e.g., cylinder or torus), $W_{\mathcal{C}}$ for non-contractible loops $\mathcal C$ are useful to characterize the topological order, as we shall see below.

The exact solution of the Hamiltonian (\ref{eq:Hamiltonian}) follows from the Majorana-fermion representation of the $\Gamma$ matrices~\cite{wen2003b,yao2009,wu2009,ryu2009},
\begin{equation}
    \Gamma^{\alpha}_j = ib^\alpha_j c_j, \quad \Gamma^{\alpha \beta}_j = ib^\alpha_j b^\beta_j,
    \label{eq:majorana}
\end{equation}
where $b^\alpha_j$ ($\alpha=1,\ldots,2q+3$) and $c_j$ are $2q+4$ Majorana operators at site $j$.
However, $2q+4$ Majorana fermions at each site span a $2^{q+2}$-dimensional Hilbert space, which is twice as large as the physical Hilbert space. To single out the physical subspace, the on-site fermion parity has to be fixed by imposing a local constraint
\begin{equation}
    D_j \equiv i^{q+2} b^1_j b^2_j \cdots b^{2q+3}_j c_j = -1,
\label{eq:constraint}
\end{equation}
which is consistent with $i^{q+1}\Gamma^1 \Gamma^2 \cdots \Gamma^{2q+3} = -1$.

By using the Majorana representation, the Hamiltonian (\ref{eq:Hamiltonian}) is rewritten as
\begin{equation}
    \tilde{H}  = \sum_{\langle ij\rangle_{\gamma}} J_{\gamma} u_{ij}
    \left(ic_i c_j + \sum_{\beta=\gamma_{\textrm{m}}+1}^{2q+3} ib_i^{\beta} b_j^{\beta}\right),
\label{eq:MajoranaHamil}
\end{equation}
where $u_{ij}=ib_i^{\gamma}b_{j}^{\gamma}$.
The solvability of $\tilde{H}$ follows the same line as the $\nu = 1$ model by observing that $[\tilde{H},u_{ij}]=[u_{ij},u_{kl}]=0$ for all different links, which divides the (enlarged) Hilbert space into subspaces with $u_{ij}$'s fixed to their eigenvalues $\pm 1$. Hence, $\tilde{H}$ describes $\nu$ species of itinerant Majorana fermions $c$ and $b^{\beta}$ in the background of a static $\mathbb{Z}_2$ gauge field, defined by the link variables $u_{ij}$. The invariance of $\tilde{H}$ under the rotation among different itinerant Majorana species is inherited from the SO($\nu$) symmetry of the Hamiltonian (\ref{eq:Hamiltonian}).

Within each subspace labeled by $u_{ij} = \pm 1$, the Hamiltonian (\ref{eq:MajoranaHamil}) becomes quadratic in itinerant Majorana fermions and can be easily diagonalized. The eigenstates of $\tilde{H}$ are generally written as $|\Psi_F(\{u\})\rangle \otimes |\{u\}\rangle$, where $|\{u\}\rangle$ refers to the static $\mathbb{Z}_2$ gauge-field configurations and $|\Psi_F(\{u\})\rangle$ are fermionic eigenstates of the quadratic Hamiltonian with fixed $u_{ij}$'s. However, not every eigenstate obtained as such satisfies the constraint in Eq.~(\ref{eq:constraint}), as is required for being a physical eigenstate of the original Hamiltonian (\ref{eq:Hamiltonian}). This, however, is easily remedied by an additional projection removing unphysical states~\cite{appendix}
\begin{equation}
    |\Psi\rangle  = P |\Psi_F(\{u\})\rangle \otimes |\{u\}\rangle,
\label{eq:eigenstate}
\end{equation}
where the projector $P = \prod_{j}(1-D_j)/2$ rigorously enforces the local constraint (\ref{eq:constraint}). Under the projection, different gauge-field configurations could result in the same wave function. To distinguish between different states, the eigenvalues of the gauge-invariant loop operators provide a useful label
\begin{equation}
    W_{\mathcal{C}} |\Psi\rangle = \prod_{\langle ij \rangle \in \mathcal{C}} u_{ij} |\Psi\rangle,
\label{eq:Wilson-loop}
\end{equation}
where the direction on each link is chosen such that $i$ ($j$) belongs to the A (B) sublattice (see Fig.~\ref{fig:lattice}).

According to Lieb's theorem~\cite{lieb1994}, the ground states of the Hamiltonian (\ref{eq:MajoranaHamil}) on the honeycomb (square) lattice appear in the zero-flux ($\pi$-flux) sector, with the eigenvalue of the elementary loops $W_p$ being $\prod_{\langle ij \rangle \in p} u_{ij} = 1$ ($-1$) following the definition in Eq.~(\ref{eq:Wilson-loop}). From now on, this is referred to as the ground-state flux configuration $\{u_0\}$. To achieve such flux choices, one may choose $u_{ij} = 1$ for all links, except for $u_{ij} = -1$ on the $4$-links of the square lattice, with the convention that $i$ ($j$) belongs to the A (B) sublattice. After fixing the gauge field, the itinerant Majorana fermions decouple from each other, and the dispersion relations for the fermionic excitations are the same as the $\nu=1$ and $\nu=2$ models on the honeycomb and square lattices, respectively. The phase diagrams as a function of $J_{\gamma}$ are hence identical to these two simplest cases~\cite{kitaev2006b,nakai2012}: Gapped phases are stabilized if one of the three (four) $|J_{\gamma}|$'s is greater than the sum of the remaining two (three) on the honeycomb (square) lattice, and a gapless phase appears otherwise.

For our purpose, we shall concentrate on the gapless phase and set $J_{\gamma} = 1$ to simplify the discussion.
For the honeycomb (square) lattice, the spectrum of an itinerant Majorana species in the zero-flux ($\pi$-flux) sector is restricted to half of the lattice's first Brillouin zone and features a single gapless Dirac cone at $\mathbf{k} = (4\pi/3,0)$ [$\mathbf{k} = (\pi/\sqrt{2},0)$]. In order to obtain models realizing the sixteenfold way, we require that these Dirac cones are gapped out and the itinerant Majorana fermions remain decoupled, each of which becomes a ``weak pairing'' topological superconductor with Chern number $1$. To that end, we add the following three-site interactions as a perturbation to the Hamiltonian (\ref{eq:Hamiltonian}):
\begin{equation}
    H'= -\kappa \! \sum_{\circlearrowright {\langle ijk \rangle}_{\gamma \gamma'} } \! \left( \Gamma^{\gamma}_i \Gamma^{\gamma \gamma'}_j \Gamma^{\gamma'}_k
    - \sum_{\beta=\gamma_{\textrm{m}}+1}^{2q+3} \! \Gamma^{\beta \gamma}_i \Gamma^{\gamma \gamma'}_j \Gamma^{\gamma' \beta}_k \right),
\label{eq:threebody}
\end{equation}
where $\circlearrowright \langle ijk \rangle_{\gamma \gamma'}$ refers to the clockwise summation over three sites within the same plaquette such that $i$ and $j$ ($j$ and $k$) are connected via a link of type $\gamma$ ($\gamma'$).

Importantly, these terms commute with the plaquette operators $W_p$ such that these remain local integrals of motion, and thus $H'$ does not mix flux sectors, in contrast to more generic perturbations~\cite{song16}.

Employing the Majorana representation~\eqref{eq:majorana}, the perturbation $H'$ is seen to give rise to next-nearest-neighbor (NNN) hopping of the itinerant Majoranas coupled to the $\mathbb{Z}_2$ gauge field,
\begin{equation}
    \tilde{H}' = \kappa \sum_{\circlearrowright {\langle ijk \rangle}_{\gamma \gamma'} } u_{ij} u_{jk} \left( i c_i c_k
    + \sum_{\beta=\gamma_{\textrm{m}}+1}^{2q+3} i b^\beta_i b^\beta_k \right).
\end{equation}
From $[u_{ij},u_{kl}]=0$ it follows that in a given gauge field configuration,
$\tilde{H}+\tilde{H}'$ corresponds to $\nu$ copies of a free Majorana hopping problem on the honeycomb or square lattice, respectively.
Fixing the ground-state flux configuration, the perturbation induces \emph{chiral} NNN hopping on the square and honeycomb lattices~\cite{appendix}.
While the Dirac cones at $\kappa = 0$ are protected by time-reversal symmetry $\mathcal{T}$ and particle-hole symmetry $\mathcal{P}$, the chiral hopping at any finite $\kappa$ breaks the time-reversal symmetry and opens up a spectral gap. The respective Majorana hopping model has $\mathcal{P}^2=+1$ and thus belongs to Class D in the free-fermion classification~\cite{altland1997,kitaev2009,ryu2010}, which is characterized by a Chern number $\nu \in \mathbb{Z}$~\cite{kitaev2006b}. As each of the $2q+4-\gamma_{\textrm{m}}$ identical Majorana hopping problems gives rise to a chiral topological superconductor with a Chern number $\mathrm{sgn}(\kappa)$~\cite{appendix}, the additivity of the topological invariant implies that the perturbed system $\tilde{H}+\tilde{H}'$ has a Chern number $\nu = 2q$ ($\nu = 2q+1$) on the square (honeycomb) lattice for $\kappa>0$.

{\em Characterizing topological order.}---The exact solvability allows us to study the sixteenfold way directly from the microscopic model, thus complementing Kitaev's axiomatic approach~\cite{kitaev2006b} based on topological quantum field theory. To establish that the microscopic models indeed provide lattice realizations of the sixteenfold way, we characterize the topological order by showing two sharp features: (i) The topological spin of an anyonic quasiparticle is $\theta = \nu\pi/8$ and (ii) the ground-state degeneracy on the torus is four (three) for even (odd) $\nu$.

\begin{figure}
\includegraphics[width=\linewidth]{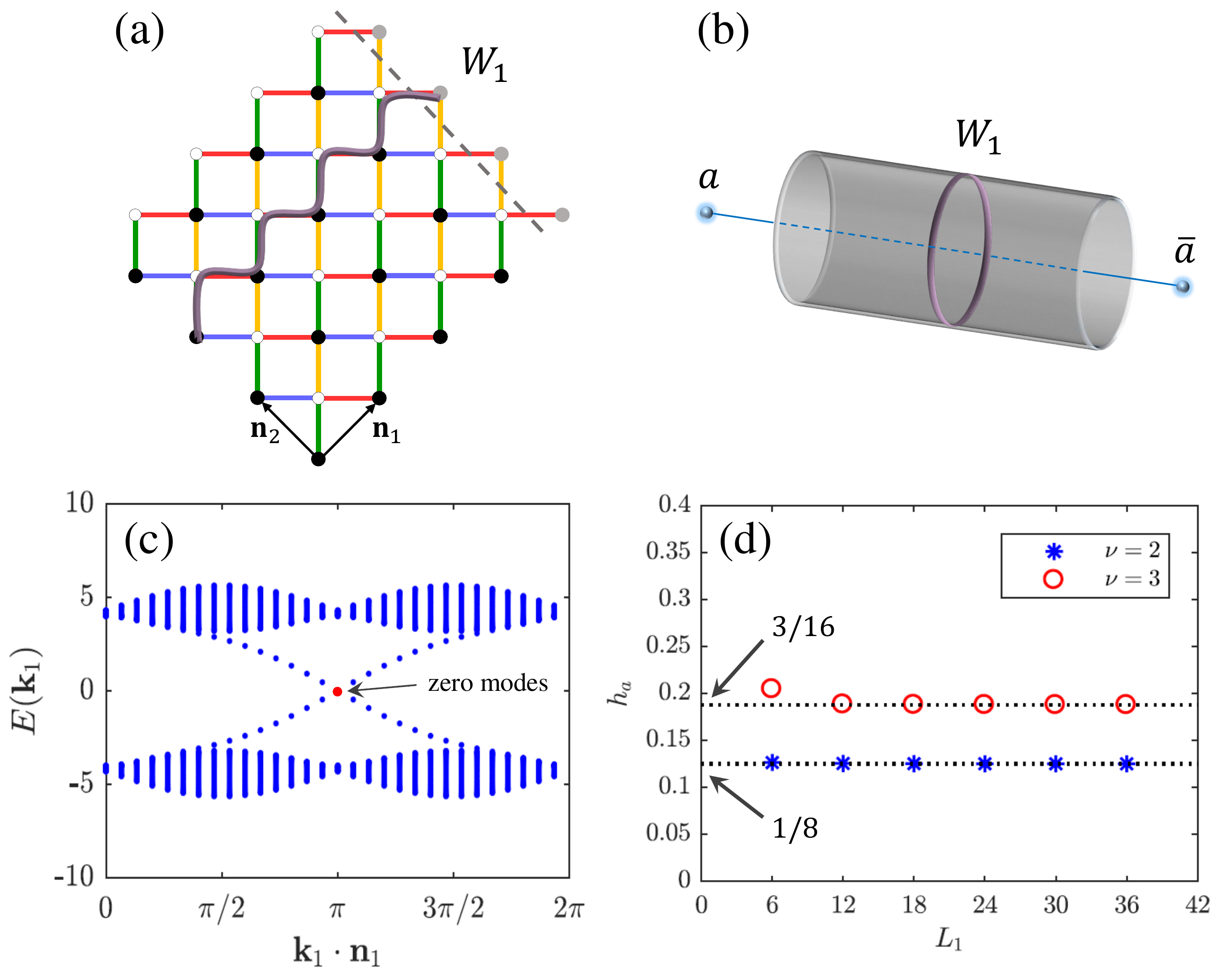}
\caption{(a) Square lattice defined on a cylinder with operator $W_1$ for a non-contractible loop. (b) Ground state with an anyon flux $a$ threading the cylinder. (c) Spectrum of the itinerant Majorana fermions for the square lattice model on the cylinder with ground-state gauge configuration and periodic boundary condition ($W_1=1$), where the red dot indicates the Majorana zero modes. (d) The extracted conformal weight of the CFT primary field vs the cylinder circumference $L_1$ for $\nu=2$ (blue stars) and $\nu=3$ (red circles) models, where the horizontal dotted lines correspond to the expected values.
The calculations in (c) and (d) are performed for cylinders of length $L_2 = 30$ [and the circumference $L_1 = L_2$ in (c)] and employ $\kappa=0.2$.}
\label{fig:cylinder}
\end{figure}

For calculating the topological spin, the lattice system is embedded on a finite cylinder with $L_1$ ($L_2$) unit cells along the $\mathbf{n}_1$ ($\mathbf{n}_2$) direction with a periodic (open) boundary, where $L_1$ and $L_2$ are taken to be even. Due to the nontrivial topology, a loop operator $W_1$ wrapping around the cylinder can be defined [see Fig.~\ref{fig:cylinder}(a) for the square lattice example]. This loop operator has eigenvalues $W_1 = \pm 1$, both of which are realizable in the ground-state flux configuration: $W_1=1$ is obtained with the aforementioned gauge choice, while $W_1 = -1$ is achieved by flipping the signs of all $u_{ij}$'s along an open path [indicated by a dashed line in Fig.~\ref{fig:cylinder}(a)]. These two choices lead to periodic boundary conditions (PBCs) and antiperiodic boundary conditions (APBCs) for itinerant Majorana fermions along the $\mathbf{n}_1$ direction, respectively.

Although the ground states in two sectors are degenerate in the thermodynamic limit, the $W_1 = -1$ sector has slightly lower energy for a finite cylinder and is hence associated with the identity sector in the anyon context. For obtaining the ground state in the $W_1 = 1$ sector, a pair of anyons (denoted by $a$ and its conjugate $\bar{a}$) are created at one boundary, and $\bar{a}$ is sent to the other boundary through the cylinder [see Fig.~\ref{fig:cylinder}(b)]. This procedure creates a state with a definite anyon flux (labeled by $a$) threading the cylinder and changes the boundary condition of itinerant Majorana fermions from APBCs to PBCs, which gives rise to $\nu$ Majorana zero modes at each boundary, as shown in Fig.~\ref{fig:cylinder}(c).

For a finite cylinder, the energy difference between the two sectors, denoted by $(E_a - E_0)$, allows us to extract the topological spin of the anyon. As the gapped bulk contributions should cancel each other for large cylinders, the energy difference receives a nontrivial contribution from the gapless edge states described by CFT~\cite{francesco1997}, $E_a - E_0 = \frac{2\pi v}{L_1} (h_a + h_{\bar{a}})$, where $a$ is now also a label for a CFT primary field ($\bar{a}$ being its conjugate), $h_a$ is the conformal weight of this field (with $h_a = h_{\bar{a}}$), and $v$ is velocity of the CFT. The topological spin of the anyon is related to $h_a$ via $\theta = 2\pi h_a$~\cite{tu2013b}. From the microscopic models, the velocity $v$ can be extracted from the linear dispersion of the edge spectrum [see, e.g., Fig.~\ref{fig:cylinder}(c)]. By calculating the energy difference, the conformal dimensions for $\nu=2$ and $\nu=3$ models are in excellent agreement with $h_a = \nu/16$ [see Fig.~\ref{fig:cylinder}(d)], which implies that the topological spin of the anyon is $\theta=\nu\pi/8$.

For calculating the ground-state degeneracy, we turn to the torus geometry by closing the boundary in the $\mathbf{n}_2$ direction, where another loop operator $W_2$ wrapping around the torus along this $\mathbf{n}_2$ direction can be defined. The two loop operators mutually commute, both having eigenvalues $\pm 1$, and lead to four sectors. The gauge choices for the four sectors are directly generalizable from the cylinder case and lead to four possible boundary conditions for the itinerant Majorana fermions, i.e., PBCs and APBCs in the $\mathbf{n}_1$ and $\mathbf{n}_2$ directions. The four candidate ground states are hence written as
\begin{equation}
    |\Psi_{\pm\pm}\rangle  = P |\Psi_F(\{u^{\pm\pm}_0\})\rangle \otimes |\{u^{\pm\pm}_0\}\rangle,
\label{eq:torus-gs}
\end{equation}
where $\pm\pm$ indicates the boundary conditions ($+$ for PBC and $-$ for APBC) of itinerant Majorana fermions in two directions and $|\Psi_F(\{u^{\pm\pm}_0\})\rangle$ is the ground state of the itinerant Majorana fermions under the respective boundary conditions. However, further analysis~\cite{appendix} reveals that for \emph{odd} $\nu$, $|\Psi_F(\{u^{++}_0\})\rangle \otimes |\{u^{++}_0\}\rangle$ has an incompatible fermion parity with the local constraint in Eq.~(\ref{eq:constraint}) and hence does not survive after projection. Thus, $|\Psi_{++}\rangle=0$ for odd $\nu$, which proves that the ground-state degeneracy is four (three) for even (odd) $\nu$ models. This agrees with even (odd) $\nu$ theories in the sixteenfold way having four (three) types of anyonic quasiparticles~\cite{kitaev2006b}.

{\em Spin-orbital realization.}---For specific values of $\nu$, we recover known models of potential relevance to Mott insulators with spin and orbital degrees of freedom by choosing a suitable representation of the $\Gamma$ matrices.

Consider the four-dimensional representation $(\Gamma^\alpha)_{\alpha=1,\dots,5}=(\sigma^y \otimes \tau^x, \sigma^y \otimes \tau^y, \sigma^y \otimes \tau^z, \sigma^x \otimes \mathbbm{1}_2, \sigma^z \otimes \mathbbm{1}_2)$, where the Pauli $\sigma$ and $\tau$ matrices are assumed to act on spin and orbital degrees of freedom, respectively. The Hamiltonian (\ref{eq:Hamiltonian}) for $\nu=2$ and $\nu=3$ can then be written as~\cite{appendix}
\begin{equation}
    H = -\sum_{\langle i j \rangle_\gamma} J_\gamma (\vec \sigma_i \cdot \vec \sigma_j) \otimes (\tau^\gamma_i \tau^\gamma_j),
\label{eq:spin-orbital}
\end{equation}
where we have abbreviated $\vec \sigma \equiv (\sigma^x, \sigma^y)$, $(\tau^\gamma)_{\gamma=1,\dots,4} = (\tau^x, \tau^y, \tau^z, \mathbbm{1})$ for $\nu=2$ and $\vec \sigma \equiv (\sigma^x, \sigma^y, \sigma^z)$, $(\tau^\gamma)_{\gamma=1,2,3} = (\tau^x, \tau^y, \tau^z)$ for $\nu = 3$, respectively. This defines a spin-orbital model on the square (honeycomb) lattice with an XY (Heisenberg) coupling in the spin sector and a Kitaev coupling in the orbital sector for $\nu=2$ ($\nu=3$). Such bond-dependent exchange interactions have been discussed previously in the context of Kugel-Khomskii-type models for transition metal oxides with strong spin-orbit coupling~\cite{natori2016,natori2019,romhanyi2017}, and belong to the larger class of compass interactions~\cite{nussinov2015}. In a real material, additional interactions will be present, spoiling exact solvability. However, the topological nature of the quantum spin-orbital liquid guarantees its stability towards arbitrary weak perturbations. In particular, the chiral edge modes lead to a $\nu/2$-quantized thermal Hall conductivity $(1/T)\kappa_{xy} = \frac{\nu}{2} [(\pi k_\mathrm{B}^2)/(6 \hbar)]$ \cite{kane1997,cappelli2002,kitaev2006b}, which is a characteristic signature of the topological ground state~\cite{vinkleraviv2018,ye2018}.

{\em Discussion.}---We have provided a systematic and complete construction of microscopic models realizing the sixteenfold way of anyon theories predicted by Kitaev. These are exactly solvable models defined using $\Gamma$ matrices satisfying the Clifford algebra, and solved in terms of a Majorana-fermion representation. Based on the exact solution, the topological order is characterized by calculating the topological spin of an anyonic quasiparticle and the ground-state degeneracy on the torus. The possible relevance of the $\nu=2$ and $\nu=3$ models to spin-orbital systems is made explicit by choosing a suitable $\Gamma$-matrix representation. It would be very interesting to see whether some of these Abelian and non-Abelian chiral spin liquids can be experimentally realized in spin-orbital materials.

{\em Acknowledgments.}---We are grateful to David Aasen, Jan Carl Budich, Meng Cheng, Tobias Meng, Qiang-Hua Wang, and Yi Zhou for helpful discussions. This research has been funded by the Deutsche Forschungsgemeinschaft (DFG) through SFB 1143 (project id 247310070), the W\"{u}rzburg-Dresden Cluster of Excellence {\it ct.qmat} (EXC 2147, project id 390858490), and the Emmy Noether program (JA2306/4-1, project id 411750675), as well as by the IMPRS for Many Particle Systems in Structured Environment at MPI-PKS.

\bibliography{Anyon}

\end{document}

\title{Supplemental Material for ``Microscopic models for Kitaev's sixteenfold way of anyon theories"}

\author{Sreejith Chulliparambil}
\affiliation{Institut f\"ur Theoretische Physik and W\"urzburg-Dresden Cluster of Excellence ct.qmat, Technische Universit\"at Dresden, 01062 Dresden, Germany}
\affiliation{Max-Planck-Institut f{\"u}r Physik komplexer Systeme, N{\"o}thnitzer Stra{\ss }e 38, 01187 Dresden, Germany}

\author{Urban F. P. Seifert}
\affiliation{Institut f\"ur Theoretische Physik and W\"urzburg-Dresden Cluster of Excellence ct.qmat, Technische Universit\"at Dresden, 01062 Dresden, Germany}

\author{Matthias Vojta}
\affiliation{Institut f\"ur Theoretische Physik and W\"urzburg-Dresden Cluster of Excellence ct.qmat, Technische Universit\"at Dresden, 01062 Dresden, Germany}

\author{Lukas Janssen}
\affiliation{Institut f\"ur Theoretische Physik and W\"urzburg-Dresden Cluster of Excellence ct.qmat, Technische Universit\"at Dresden, 01062 Dresden, Germany}

\author{Hong-Hao Tu}
\email{hong-hao.tu@tu-dresden.de}
\affiliation{Institut f\"ur Theoretische Physik and W\"urzburg-Dresden Cluster of Excellence ct.qmat, Technische Universit\"at Dresden, 01062 Dresden, Germany}

\begin{abstract}
This Supplemental Material provides more details about the diagonalization of the quadratic Hamiltonians for itinerant Majorana fermions, the calculation of the ground-state degeneracy on the torus, and the SO($\nu$) symmetry of the microscopic Hamiltonian.
\end{abstract}

\maketitle
\tableofcontents

\setcounter{figure}{0}
\setcounter{equation}{0}
\renewcommand\thefigure{S\arabic{figure}}
\renewcommand\theequation{S\arabic{equation}}

\section{Diagonalizing quadratic Hamiltonians of Majorana fermions}

In this section, we diagonalize the quadratic Hamiltonians of itinerant Majorana fermions, defined by
\begin{equation}
\tilde{H}=\sum_{\langle ij\rangle _{\gamma }}J_{\gamma }u_{ij}\left(
ic_{i}c_{j}+\sum_{\beta =\gamma _{\text{m}}+1}^{2q+3}ib_{i}^{\beta
}b_{j}^{\beta }\right),  \label{eq:H1}
\end{equation}%
with coupling $J_{\gamma }=1$ for all $\gamma = 1,\dots,\gamma_\text{m}$, $\gamma_\text{m} = 4$ (3) on the square (honeycomb) lattice, and the perturbation%
\begin{equation}
\tilde{H}^{\prime }=\kappa \sum_{\circlearrowright {\langle ijk\rangle }%
_{\gamma \gamma ^{\prime }}}u_{ij}u_{jk}\left( ic_{i}c_{k}+\sum_{\beta
=\gamma _{\text{m}}+1}^{2q+3}ib_{i}^{\beta }b_{k}^{\beta }\right) .
\label{eq:H2}
\end{equation}%
Here we restrict ourselves to the ground-state flux configuration, i.e., $\pi $-flux (zero-flux) for the square (honeycomb) lattice. Once the gauge choice for $u_{ij}$'s is fixed, the itinerant Majorana fermions decouple and it is sufficient to consider a single itinerant Majorana species, for which we choose $c$-Majorana fermion below.

\subsection{Square-lattice model}

For the square-lattice model, the gauge choice realizing the $\pi $-flux can
be chosen as%
\begin{equation}
u_{ij}=\left\{
\begin{array}{c}
1 \\
-1%
\end{array}%
\right. \left.
\begin{array}{c}
\text{for }\gamma =1,2,3 \\
\text{for }\gamma =4%
\end{array}%
\right. ,  \label{eq:pi-flux}
\end{equation}%
where we have adopted the convention that $i$ and $j$ belong to A and B
sublattices, respectively.

\begin{figure}[tbp]
\includegraphics[width=0.65\textwidth]{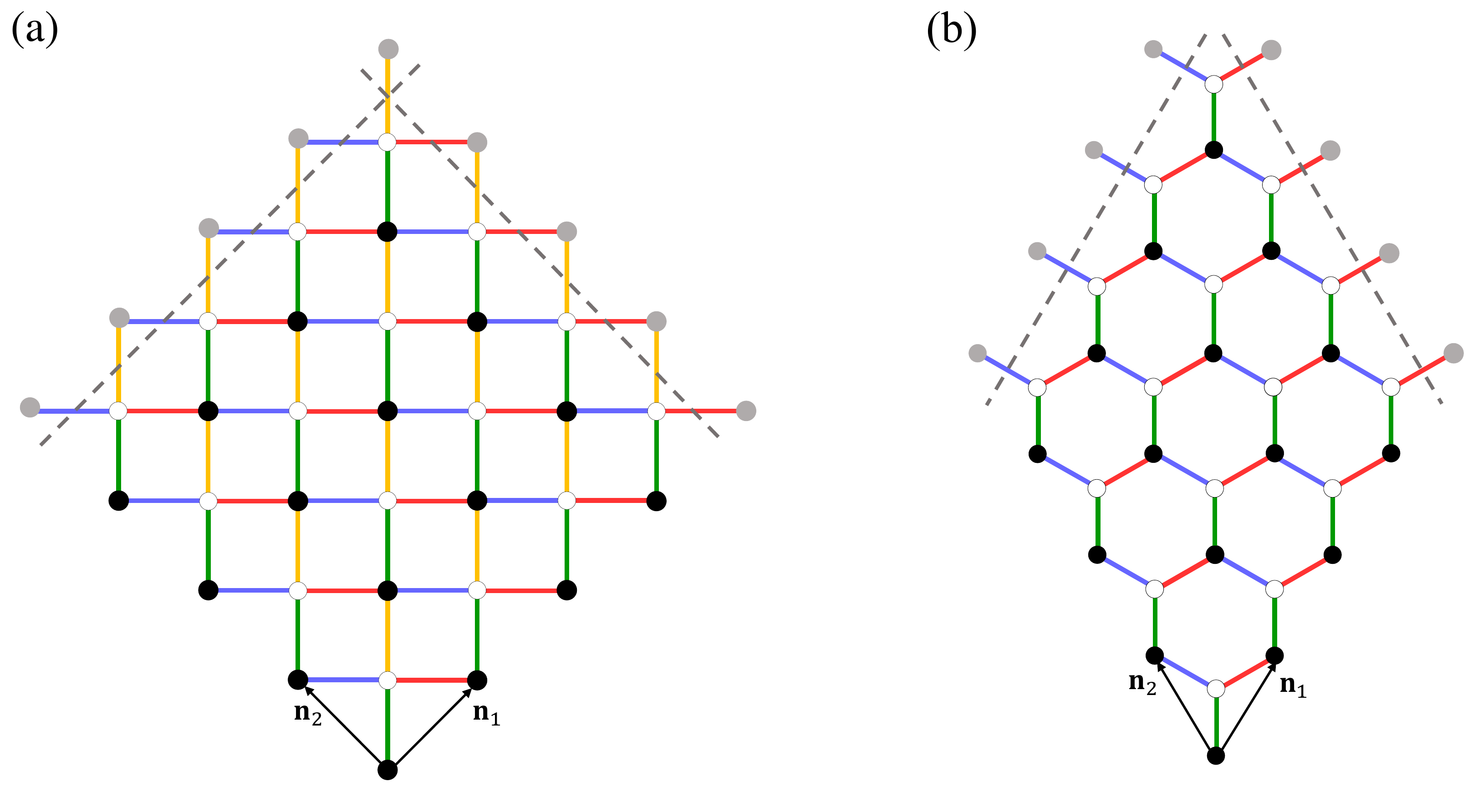}
\centering
\caption{(a) Square and (b) honeycomb lattices defined on the torus. The periodic or antiperiodic boundary condition of the itinerant Majorana fermions along $\mathbf{n}_1$ and $\mathbf{n}_2$ directions are indicated by whether the signs of $u_{ij}$'s on the dashed lines are flipped.}
\label{fig:torus}
\end{figure}

We define two sites connected with 3-links (green links in Fig.~\ref{fig:torus}) as a unit cell and label the two Majorana fermions within the same unit cell by a sublattice index, i.e., $c_{j,s}$ with $s=\text{A},\text{B}$. The Fourier transform for the Majorana fermions is then written as
\begin{equation}
c_{j,s}=\sqrt{\frac{2}{N}}\sum_{\mathbf{k}}c_{\mathbf{k},s}e^{i\mathbf{k}
\cdot \mathbf{R}_{j}},  \label{eq:Fourier}
\end{equation}
where the coordinate for the A sublattice (black dots in Fig.~\ref{fig:torus}) is chosen to define the lattice vector $\mathbf{R}_{j}$ for the $j$-th unit cell, $\mathbf{k}$ belongs to the first Brillouin zone (BZ), and $N$ is the number of unit cells. The inverse Fourier transform is given by
\begin{equation}
c_{\mathbf{k},s}=\sqrt{\frac{1}{2N}}\sum_{j}c_{j,s}e^{-i\mathbf{k}\cdot
\mathbf{R}_{j}},  \label{eq:invFourier}
\end{equation}%
which satisfies $c_{-\mathbf{k},s}=c_{\mathbf{k},s}^{\dagger }$. Thus, it is convenient to consider $\mathbf{k}$ points in half of the first BZ (denoted by $\mathbf{k}\in \text{BZ}/2$) and use the anticommutation relation $\{c_{\mathbf{k},s},c_{\mathbf{k}^{\prime },s^{\prime }}^{\dagger }\}=\delta _{\mathbf{kk}^{\prime }}\delta _{ss^{\prime }}$, where $c_{\mathbf{k},s}^{\dagger }$ and $c_{\mathbf{k},s}$ can be viewed as ordinary fermionic creation and annihilation operators, respectively. We note, however, that the so-called time-reversal (TR) invariant $\mathbf{k}$ points satisfying $\mathbf{k}=\mathbf{-k}+\mathbf{G}$ ($\mathbf{G}$: reciprocal lattice vectors), if they exist, have to be treated with special care. For the TR-invariant points, one has $c_{-\mathbf{k},s}=c_{\mathbf{k},s}^{\dagger }=c_{\mathbf{k},s}$ and $(c_{\mathbf{k},s})^{2}=1/2$, so it could be rescaled to define a Majorana mode $\tilde{c}_{\mathbf{k},s}=\sqrt{2}c_{\mathbf{k},s}$ satisfying $(\tilde{c}_{\mathbf{k},s})^{2}=1$. This will be of particular importance for the discussions in the next section. Since the thermodynamic limit will be taken, we will not pay special attention to these TR invariant points in the present section.

By using the Fourier transform (\ref{eq:Fourier}), the $c$-Majorana Hamiltonian from Eq.~(\ref{eq:H1}), under the gauge choice (\ref{eq:pi-flux}), is written as
\begin{eqnarray}
\tilde{H} &=&2i\sum_{\mathbf{k}}[e^{-i\mathbf{k}\cdot \mathbf{n}_{1}}+e^{-i\mathbf{k}\cdot \mathbf{n}_{2}}+1-e^{-i\mathbf{k}\cdot (\mathbf{n}_{1}+
\mathbf{n}_{2})}]c_{-\mathbf{k},\text{A}}c_{\mathbf{k},\text{B}}  \notag \\
&=&\sum_{\mathbf{k}\in \text{BZ}/2}
\begin{pmatrix}
c_{\mathbf{k},\text{A}}^{\dagger } & c_{\mathbf{k},\text{B}}^{\dagger}
\end{pmatrix}
\begin{pmatrix}
0 & if(\mathbf{k}) \\
-if^{\ast }(\mathbf{k}) & 0
\end{pmatrix}
\begin{pmatrix}
c_{\mathbf{k},\text{A}} \\
c_{\mathbf{k},\text{B}}
\end{pmatrix}
\label{eq:squareH1}
\end{eqnarray}
with
\begin{equation}
f(\mathbf{k})=2[e^{-i\mathbf{k}\cdot \mathbf{n}_{1}}+e^{-i\mathbf{k}\cdot
\mathbf{n}_{2}}+1-e^{-i\mathbf{k}\cdot (\mathbf{n}_{1}+\mathbf{n}_{2})}].
\label{eq:f-function}
\end{equation}
Thus, the single-particle excitation energy is given by $\varepsilon (\mathbf{k})=\pm |f(\mathbf{k})|$ with $\mathbf{k}\in \text{BZ}/2$, which has a Dirac cone at $\mathbf{k}^{\ast }=(\pi /\sqrt{2},0)$.

Similarly, the $c$-Majorana Hamiltonian from Eq.~(\ref{eq:H2}) with the gauge choice (\ref{eq:pi-flux}) is given by
\begin{eqnarray}
\tilde{H}^{\prime } &=&-4i\kappa \sum_{\mathbf{k}}(e^{i\mathbf{k}\cdot
\mathbf{n}_{1}}+e^{-i\mathbf{k}\cdot \mathbf{n}_{2}})c_{-\mathbf{k},\text{A}}c_{\mathbf{k},\text{A}}+4i\kappa \sum_{\mathbf{k}}(e^{i\mathbf{k}\cdot
\mathbf{n}_{1}}+e^{-i\mathbf{k}\cdot \mathbf{n}_{2}})c_{-\mathbf{k},\text{B}}c_{\mathbf{k},\text{B}}  \notag \\
&=&\sum_{\mathbf{k}\in \mathrm{BZ}/2}%
\begin{pmatrix}
c_{\mathbf{k},\text{A}}^{\dagger } & c_{\mathbf{k},\text{B}}^{\dagger }
\end{pmatrix}
\begin{pmatrix}
\Delta (\mathbf{k}) & 0 \\
0 & -\Delta (\mathbf{k})%
\end{pmatrix}%
\begin{pmatrix}
c_{\mathbf{k},\text{A}} \\
c_{\mathbf{k},\text{B}}%
\end{pmatrix}%
\label{eq:squareH2}
\end{eqnarray}%
with
\begin{equation}
\Delta (\mathbf{k})=8\kappa \lbrack \sin (\mathbf{k}\cdot \mathbf{n}
_{1})-\sin (\mathbf{k}\cdot \mathbf{n}_{2})].  \label{eq:Delta-function}
\end{equation}

By combining the Hamiltonian terms in Eq.~(\ref{eq:squareH1}) and (\ref{eq:squareH2}), we obtain
\begin{equation}
\tilde{H}+\tilde{H}^{\prime }=\sum_{\mathbf{k}\in \mathrm{BZ}/2}%
\begin{pmatrix}
c_{\mathbf{k},\text{A}}^{\dagger } & c_{\mathbf{k},\text{B}}^{\dagger }%
\end{pmatrix}%
\begin{pmatrix}
\Delta (\mathbf{k}) & if(\mathbf{k}) \\
-if^{\ast }(\mathbf{k}) & -\Delta (\mathbf{k})%
\end{pmatrix}%
\begin{pmatrix}
c_{\mathbf{k},\text{A}} \\
c_{\mathbf{k},\text{B}}%
\end{pmatrix}
\label{eq:totalH}
\end{equation}%
with single-particle excitation energy $E^{\pm }(\mathbf{k})=\pm \sqrt{|f(\mathbf{k})|^{2}+\Delta (\mathbf{k})^{2}}$, where the Dirac point at $\mathbf{k}^{\ast }=(\pi /\sqrt{2},0)$ is gapped out since $\Delta (\mathbf{k}^{\ast })=16\kappa $.

Now we calculate the Chern number by writting the Hamiltonian as
\begin{equation}
\tilde{H}+\tilde{H}^{\prime }=\sum_{\mathbf{k}\in \mathrm{BZ}/2}c_{\mathbf{k}%
}^{\dagger }H(\mathbf{k})c_{\mathbf{k}},
\label{eq:totalH-compact}
\end{equation}
where $c_{\mathbf{k}}=
\begin{pmatrix}
c_{\mathbf{k},\text{A}} & c_{\mathbf{k},\text{B}}
\end{pmatrix}%
^{T}$. Here we have defined $H(\mathbf{k})=\vec{h}(\mathbf{k})\cdot \vec{\sigma}$ with $\vec{\sigma}=(\sigma ^{x},\sigma ^{y},\sigma ^{z})$ and $\vec{h}(\mathbf{k})=(h_{x}(\mathbf{k}),h_{y}(\mathbf{k}),h_{z}(\mathbf{k}))$, where $h_{x}(\mathbf{k})=-$Im$f(\mathbf{k})$, $h_{y}(\mathbf{k})=-$Re$f(\mathbf{k})$, and $h_{z}(\mathbf{k})=\Delta (\mathbf{k})$. With this form, the Chern number is defined by
\begin{equation}
C=\frac{1}{4\pi }\int_{\mathrm{BZ}}d^{2}\mathbf{k}\text{ }\mathbf{\hat{h}}
\cdot (\partial _{k_{x}}\mathbf{\hat{h}}\times \partial _{k_{y}}\mathbf{\hat{
h}})  \label{eq:Chern-number}
\end{equation}
with unit vector $\mathbf{\hat{h}}=\vec{h}(\mathbf{k})/E^{+}(\mathbf{k})$. By using the explicit form of $f(\mathbf{k})$ and $\Delta (\mathbf{k})$ in Eqs.~(\ref{eq:f-function}) and (\ref{eq:Delta-function}), we obtain
\begin{equation}
C=\mathrm{sgn}(\kappa).
\end{equation}

\subsection{Honeycomb-lattice model}

For the honeycomb-lattice model, the gauge choice realizing the zero-flux is chosen as
\begin{equation}
u_{ij}=1\ \text{for }\gamma =1,2,3,  \label{eq:zero-flux}
\end{equation}%
with the convention that $i$ and $j$ belong to A and B sublattices, respectively.

Similar to the above square-lattice model, two sites connected with 3-links are defined as a unit cell for the honeycomb lattice. The procedure of diagonalizing the quadratic Hamiltonian of $c$-Majorana fermion follows the same steps. In momentum space, the Hamiltonian $\tilde{H}+\tilde{H}^{\prime}$ takes an identical form as Eq.~(\ref{eq:totalH}), except that $f(\mathbf{k})$ and $\Delta (\mathbf{k})$ for the honeycomb-lattice model are given by
\begin{eqnarray}
f(\mathbf{k}) &=&2(e^{-i\mathbf{k}\cdot \mathbf{n}_{1}}+e^{-i\mathbf{k}\cdot
\mathbf{n}_{2}}+1), \\
\Delta (\mathbf{k}) &=&4\kappa \{\sin (\mathbf{k}\cdot \mathbf{n}_{1})-\sin (%
\mathbf{k}\cdot \mathbf{n}_{2})-\sin [\mathbf{k}\cdot (\mathbf{n}_{1}-%
\mathbf{n}_{2})]\}.
\end{eqnarray}%
For $\kappa =0$, the single-particle dispersion relation $\varepsilon(\mathbf{k})=\pm |f(\mathbf{k})|$ has a Dirac cone at $\mathbf{k}^{\ast }=(4\pi/3,0)$, which is gapped out for any finite $\kappa \neq 0$ since $\Delta (\mathbf{k}^{\ast })=6\sqrt{3}\kappa $. By using the definition of the Chern number in Eq.~(\ref{eq:Chern-number}), we also obtain $C=\mathrm{sgn}(\kappa)$ for the $c$-Majorana fermion on the honeycomb lattice.

\section{Ground-state degeneracy on the torus}

In this section, we prove that the ground-state degeneracy on the torus is four (three) on the square (honeycomb) lattice. The lattice is embedded on a finite torus with $L_{1}$ ($L_{2}$) unit cells with periodic boundary along $\mathbf{n}_{1}$ ($\mathbf{n}_{2}$) directions. We take both $L_{1}$ and $L_{2}$ to be \textit{even}.

\subsection{Definition of the fermion parity}

The key for the proof is to calculate the total fermion parity, including both itinerant Majorana fermions and Majorana fermions forming static $\mathbb{Z}_{2}$ gauge fields.

Let us start with a careful definition of the fermion parity. In the Majorana representation of the $\Gamma$ matrices, the on-site fermion parity has to fulfill the local constraint
\begin{equation}
i^{q+2}b^{1}b^{2}\ldots b^{2q+3}c=-1,
\end{equation}
where, for ease of notation, we have suppressed the site index.

For both square and honeycomb lattices, the gauge choice for the ground-state flux configuration gives rise to a two-site unit cell. Thus, we define the fermion parity operator within a unit cell as $(ib_{\text{A}}^{1}b_{\text{B}}^{1})(ib_{\text{A}}^{2}b_{\text{B}}^{2})\ldots (ib_{\text{A}}^{2q+3}b_{\text{B}}^{2q+3})(ic_{\text{A}}c_{\text{B}})$, which is actually fixed by the above on-site fermion parity
\begin{eqnarray}
(ib_{\text{A}}^{1}b_{\text{B}}^{1})(ib_{\text{A}}^{2}b_{\text{B}}^{2})\ldots
(ib_{\text{A}}^{2q+3}b_{\text{B}}^{2q+3})(ic_{\text{A}}c_{\text{B}})
&=&(-1)^{q}(i^{q+2}b_{\text{A}}^{1}b_{\text{A}}^{2}\ldots b_{\text{A}%
}^{2q+3}c_{\text{A}})(i^{q+2}b_{\text{B}}^{1}b_{\text{B}}^{2}\ldots b_{\text{%
B}}^{2q+3}c_{\text{B}})  \notag \\
&=&(-1)^{q}.
\label{eq:unitcell-parity}
\end{eqnarray}

By using the above definition of fermion parity in each unit cell, the total fermion parity for the whole lattice is defined by
\begin{equation}
Q=\prod_{j}(ib_{j,\text{A}}^{1}b_{j,\text{B}}^{1})(ib_{j,\text{A}}^{2}b_{j,\text{B}}^{2})\ldots (ib_{j,\text{A}}^{2q+3}b_{j,\text{B}}^{2q+3})(ic_{j,\text{A}}c_{j,\text{B}}).
\label{eq:total-parity}
\end{equation}
By using Eq.~(\ref{eq:unitcell-parity}), we can see that the total fermion parity must be \textit{even}, i.e., $Q=(-1)^{qN}=1$ since $N=L_{1}L_{2}$ is even.

For our models, the total fermion parity $Q$ factorizes into two parts, one for itinerant Majorana fermions and another for static $\mathbb{Z}_{2}$ gauge fields. Accordingly, we rewrite the total fermion parity defined in Eq.~(\ref{eq:total-parity}) as
\begin{equation}
Q=Q_{1}Q_{2},
\end{equation}
where $Q_{1}$ is the fermion parity for the itinerant Majorana fermions
\begin{equation}
Q_{1}=\left\{
\begin{array}{c}
\prod_{j}(ib_{j,\text{A}}^{5}b_{j,\text{B}}^{5})\cdots (ib_{j,\text{A}}^{2q+3}b_{j,\text{B}}^{2q+3})(ic_{j,\text{A}}c_{j,\text{B}}) \\
\prod_{j}(ib_{j,\text{A}}^{4}b_{j\text{B}}^{4})\cdots (ib_{j,\text{A}}^{2q+3}b_{j,\text{B}}^{2q+3})(ic_{j,\text{A}}c_{j,\text{B}})
\end{array}
\right. \left.
\begin{array}{c}
\text{for square lattice} \\
\text{for honeycomb lattice}%
\end{array}%
\right. ,
\label{eq:Q1-parity}
\end{equation}%
and $Q_{2}$ the fermion parity for static $\mathbb{Z}_{2}$ gauge fields
\begin{equation}
Q_{2}=\left\{
\begin{array}{c}
\prod_{j}(ib_{j,\text{A}}^{1}b_{j,\text{B}}^{1})\cdots (ib_{j,\text{A}}^{4}b_{j,\text{B}}^{4}) \\
\prod_{j}(ib_{j,\text{A}}^{1}b_{j,\text{B}}^{1})\cdots (ib_{j,\text{A}}^{3}b_{j,\text{B}}^{3})
\end{array}
\right. \left.
\begin{array}{c}
\text{for square lattice} \\
\text{for honeycomb lattice}
\end{array}
\right. .
\label{eq:Q2-parity}
\end{equation}

\subsection{Calculating fermion parity of itinerant Majorana fermions in momentum space}

For itinerant Majorana fermions, we calculate the fermion parity in momentum space. Below we establish a way to do such calculations. We illustrate this by using the $c$-Majorana fermion. The total fermion parity for the $c$-Majorana fermion is defined by
\begin{equation}
Q_{c}=\prod_{j}ic_{j,\text{A}}c_{j,\text{B}},
\end{equation}%
which can be represented in momentum space as follows:%
\begin{eqnarray}
Q_{c} &=&\prod_{j}(-1)^{\frac{1}{2}(1-ic_{j,\text{A}}c_{j,\text{B}})}  \notag
\\
&=&(-1)^{N/2}\exp \left( -i\pi \sum_{j}\frac{i}{2}c_{j,\text{A}}c_{j,\text{B}}\right)   \notag \\
&=&(-1)^{N/2}\exp \left( -i\pi \sum_{\mathbf{k}}ic_{-\mathbf{k},\text{A}}c_{\mathbf{k},\text{B}}\right)   \notag \\
&=&(-1)^{N/2}\exp \left( -i\pi \sum_{\mathbf{k}=\mathbf{-k}+\mathbf{G}}ic_{\mathbf{k},\text{A}}c_{\mathbf{k},\text{B}}\right) \exp \left[ -i\pi \sum_{\mathbf{k}\in \mathrm{BZ}/2}i(c_{\mathbf{k},\text{A}}^{\dagger }c_{\mathbf{k},\text{B}}-c_{\mathbf{k},\text{B}}^{\dagger}c_{\mathbf{k},\text{A}})\right]
,
\end{eqnarray}%
where $\mathbf{k}\in \mathrm{BZ}/2$ \textit{excludes} all TR-invariant points. For the TR-invariant points, we use Majorana operators $\tilde{c}_{\mathbf{k},s}=\sqrt{2}c_{\mathbf{k},s}$. For $\mathbf{k}\in \mathrm{BZ}/2$, we introduce a unitary basis rotation
\begin{equation}
\begin{pmatrix}
d_{\mathbf{k},1} \\
d_{\mathbf{k},2}%
\end{pmatrix}%
=U%
\begin{pmatrix}
c_{\mathbf{k},\text{A}} \\
c_{\mathbf{k},\text{B}}%
\end{pmatrix}%
,\text{ \ \ \ \ \ }%
\begin{pmatrix}
c_{\mathbf{k},\text{A}} \\
c_{\mathbf{k},\text{B}}%
\end{pmatrix}%
=U^{\dagger }%
\begin{pmatrix}
d_{\mathbf{k},1} \\
d_{\mathbf{k},2}%
\end{pmatrix}%
,
\label{eq:d-fermion}
\end{equation}%
where the unitary matrix $U$ is given by
\begin{equation}
U=\frac{1}{\sqrt{2}}%
\begin{pmatrix}
1 & i \\
1 & -i%
\end{pmatrix}%
.  \label{eq:unitary-rotation}
\end{equation}

After these steps, the fermion parity operator $Q_{c}$ becomes
\begin{eqnarray}
Q_{c} &=&(-1)^{N/2}\exp \left( -i\frac{\pi }{2}\sum_{\mathbf{k}=\mathbf{-k}+\mathbf{G}}i\tilde{c}_{\mathbf{k},\text{A}}\tilde{c}_{\mathbf{k},\text{B}}\right) \exp \left[-i\pi \sum_{\mathbf{k}\in \mathrm{BZ}/2}(d_{\mathbf{k},1}^{\dagger }d_{\mathbf{k},1}-d_{\mathbf{k},2}^{\dagger }d_{\mathbf{k},2})
\right]   \notag \\
&=&(-1)^{N/2}\exp \left( -i\frac{\pi }{2}\sum_{\mathbf{k}=\mathbf{-k}+\mathbf{G}}i\tilde{c}_{\mathbf{k},\text{A}}\tilde{c}_{\mathbf{k},\text{B}
}\right) (-1)^{\sum_{\mathbf{k}\in \mathrm{BZ}/2}(d_{\mathbf{k},1}^{\dagger}d_{\mathbf{k},1}+d_{\mathbf{k},2}^{\dagger }d_{\mathbf{k},2})},
\label{eq:Qc-parity}
\end{eqnarray}%
where $i\tilde{c}_{\mathbf{k},\text{A}}\tilde{c}_{\mathbf{k},\text{B}}$ can be viewed as the fermion parity for a TR-invariant momentum and the last term counts the parity of occupied $d_{\mathbf{k}}$-fermions.

\subsection{Ground-state fermion parity on the torus}

As we have shown in the main text, four (candidate) ground-state wave functions for the sixteenfold-way models defined on the torus are given by
\begin{equation}
|\Psi _{\pm \pm }\rangle =P|\Psi _{F}(\{u_{0}^{\pm \pm }\})\rangle \otimes|\{u_{0}^{\pm \pm }\}\rangle ,
\label{eq:torus-degeneracy}
\end{equation}%
where $\pm \pm $ indicates the boundary conditions [$+$ for periodic boundary condition (PBC) and $-$ for antiperiodic boundary condition (APBC)] of itinerant Majorana fermions along $\mathbf{n}_{1}$ and $\mathbf{n}_{2}$ directions and $|\Psi _{F}(\{u_{0}^{\pm \pm }\})\rangle $ is the ground state of the itinerant Majorana fermions under the respective boundary conditions. Here $\{u_{0}^{\pm \pm }\}$ is the gauge field configuration in the ground-state flux sector, for which $\{u_{0}^{++}\}$ is defined in Eq.~(\ref{eq:pi-flux}) [Eq.~(\ref{eq:zero-flux})] for the square (honeycomb) lattice, where the itinerant Majorana fermions have PBC in both directions. The remaining three gauge field configurations $\{u_{0}^{+-}\}$, $\{u_{0}^{-+}\}$, and $\{u_{0}^{--}\}$ are obtained by flipping the signs of two (closed) paths of $u_{ij}$'s (indicated by two dashed lines in Fig.~\ref{fig:torus}), thus giving rise to APBC for itinerant Majorana fermions.

The total fermion parity constraint $Q=1$ requires that
\begin{equation}
Q|\Psi _{F}(\{u_{0}^{\pm \pm }\})\rangle \otimes |\{u_{0}^{\pm \pm}\}\rangle =|\Psi _{F}(\{u_{0}^{\pm \pm }\})\rangle \otimes |\{u_{0}^{\pm
\pm }\}\rangle ,
\end{equation}%
otherwise the state cannot survive projection in Eq.~(\ref{eq:torus-degeneracy}). This requires a careful analysis of the fermion parity in $|\Psi_{F}(\{u_{0}^{\pm \pm }\})\rangle \otimes |\{u_{0}^{\pm \pm }\}\rangle $. Since $Q=Q_{1}Q_{2}$, we have
\begin{equation}
Q|\Psi _{F}(\{u_{0}^{\pm \pm }\})\rangle \otimes |\{u_{0}^{\pm \pm}\}\rangle =Q_{1}|\Psi _{F}(\{u_{0}^{\pm \pm }\})\rangle \otimes
Q_{2}|\{u_{0}^{\pm \pm }\}\rangle ,
\end{equation}
where the fermion parities for $|\Psi _{F}(\{u_{0}^{\pm \pm }\})\rangle $ and $|\{u_{0}^{\pm \pm }\}\rangle $ will be separately calculated below.

\subsubsection{Itinerant Majorana fermions}

As we have discussed above, the $\nu $ species of itinerant Majorana fermions decouple after fixing the gauge-field configuration, each of which has the \textit{same} fermion parity in $|\Psi _{F}(\{u_{0}^{\pm \pm}\})\rangle $. Thus, it is sufficient to calculate the fermion parity for the $c$-Majorana fermion and take the $\nu $-th power. For even $\nu$, it already implies that this parity is even for all four states $|\Psi _{F}(\{u_{0}^{\pm \pm}\})\rangle $. However, for odd $\nu$, a careful analysis is needed.

The subtlety of the boundary conditions for itinerant Majorana fermions is that for a finite-size torus, it determines the allowed lattice momenta in Eq.~(\ref{eq:Fourier}), which would then affect the number of TR-invariant points. We analyze individually all four boundary conditions below.

\textbf{(i) APBC in both directions (AA-type):}

In this case, the allow lattice momenta in Eq.~(\ref{eq:Fourier}) are given by $e^{i\mathbf{k}\cdot L_{1}\mathbf{n}_{1}}=e^{i\mathbf{k}\cdot L_{2}\mathbf{n}_{2}}=-1$ with
\begin{eqnarray}
\mathbf{k}\cdot \mathbf{n}_{1} &=&\pm \frac{\pi }{L_{1}},\pm \frac{3\pi }{L_{1}},\ldots ,\pm \frac{(L_{1}-1)\pi }{L_{1}},  \notag \\
\mathbf{k}\cdot \mathbf{n}_{2} &=&\pm \frac{\pi }{L_{2}},\pm \frac{3\pi }{L_{2}},\ldots ,\pm \frac{(L_{2}-1)\pi }{L_{2}},
\label{eq:APBC-APBC}
\end{eqnarray}%
where we have used that both $L_{1}$ and $L_{2}$ are even.

For the square (honeycomb) lattice, the four TR-invariant points are $\mathbf{k}=(0,0)$, $(\pm \frac{\pi }{\sqrt{2}},\frac{\pi }{\sqrt{2}})$, and $(0,\sqrt{2}\pi )$ [$\mathbf{k}=(0,0)$, $(\pm \pi ,\frac{\pi }{\sqrt{3}})$, and $(0,\frac{2}{\sqrt{3}}\pi )$]. By using the corresponding primitive vectors $\mathbf{n}_{1}$ and $\mathbf{n}_{2}$, one obtains $\mathbf{k\cdot n}_{1}=0$, $\pi $ and $\mathbf{k\cdot n}_{2}=0$, $\pi $ for the TR-invariant points. By comparing with Eq.~(\ref{eq:APBC-APBC}), we see that \textit{none} of the four TR-invariant points is allowed by AA-type boundary condition.

For calculating the $c$-Majorana fermion parity in $|\Psi_{F}(\{u_{0}^{--}\})\rangle $, we can proceed with the $c$-Majorana Hamiltonian in Eq.~(\ref{eq:totalH-compact}) without extra work on the TR-invariant points and switch to the $d_{\mathbf{k}}$-fermion basis [see Eq.~(\ref{eq:d-fermion})],
\begin{eqnarray}
\tilde{H}+\tilde{H}^{\prime } &=&\sum_{\mathbf{k}\in \mathrm{BZ}/2}c_{\mathbf{k}}^{\dagger }H(\mathbf{k})c_{\mathbf{k}}  \notag \\
&=&\sum_{\mathbf{k}\in \mathrm{BZ}/2}d_{\mathbf{k}}^{\dagger }M(\mathbf{k})d_{\mathbf{k}},
\end{eqnarray}
where $d_{\mathbf{k}}=
\begin{pmatrix}
d_{\mathbf{k},1} & d_{\mathbf{k},2}
\end{pmatrix}
^{T}$ and $M(\mathbf{k})=UH(\mathbf{k})U^{\dagger }$. Since $|\Psi_{F}(\{u_{0}^{--}\})\rangle $ is the ground state of $\tilde{H}+\tilde{H}^{\prime }$ with \textit{half-filled} $d_{\mathbf{k}}$-fermions, it is then clear that the parity of $d_{\mathbf{k}}$-fermion in $|\Psi_{F}(\{u_{0}^{--}\})\rangle $ should be
\begin{equation}
(-1)^{\sum_{\mathbf{k}\in \mathrm{BZ}/2}(d_{\mathbf{k},1}^{\dagger }d_{\mathbf{k},1}+d_{\mathbf{k},2}^{\dagger }d_{\mathbf{k},2})}|\Psi_{F}(\{u_{0}^{--}\})\rangle =(-1)^{N/2}|\Psi _{F}(\{u_{0}^{--}\})\rangle .
\end{equation}%
By using Eq.~(\ref{eq:Qc-parity}), we arrive at
\begin{eqnarray}
Q_{c}|\Psi _{F}(\{u_{0}^{--}\})\rangle  &=&(-1)^{N/2}(-1)^{\sum_{\mathbf{k}\in \mathrm{BZ}/2}(d_{\mathbf{k},1}^{\dagger }d_{\mathbf{k},1}+d_{\mathbf{k},2}^{\dagger }d_{\mathbf{k},2})}|\Psi _{F}(\{u_{0}^{--}\})\rangle   \notag \\
&=&(-1)^{N}|\Psi _{F}(\{u_{0}^{--}\})\rangle   \notag \\
&=&|\Psi _{F}(\{u_{0}^{--}\})\rangle ,
\end{eqnarray}%
which indicates that the $c$-Majorana fermion has \textit{even} fermion parity in $|\Psi _{F}(\{u_{0}^{--}\})\rangle$.

\textbf{(ii) APBC in }$n_{1}$\textbf{-direction and PBC in }$n_{2}$\textbf{-direction (AP-type):}

In this case, the allowed lattice momenta satisfy $e^{i\mathbf{k}\cdot L_{1}\mathbf{n}_{1}}=-1$ and $e^{i\mathbf{k}\cdot L_{2}\mathbf{n}_{2}}=1$ with
\begin{eqnarray}
\mathbf{k}\cdot \mathbf{n}_{1} &=&\pm \frac{\pi }{L_{1}},\pm \frac{3\pi }{L_{1}},\ldots ,\pm \frac{(L_{1}-1)\pi }{L_{1}}, \\
\mathbf{k}\cdot \mathbf{n}_{2} &=&0,\pm \frac{2\pi }{L_{2}},\pm \frac{4\pi}{L_{2}},\ldots ,\pm \frac{(L_{2}-2)\pi }{L_{2}},\pi ,
\end{eqnarray}
where \textit{none} of the four TR-invariant points is allowed. Thus, the analysis for the parity of $c$-Majorana fermion is identical to the above AA-type in case (i), which indicates that $|\Psi _{F}(\{u_{0}^{-+}\})\rangle $ also has \textit{even} parity for $c$-Majorana fermion
\begin{equation}
Q_{c}|\Psi _{F}(\{u_{0}^{-+}\})\rangle =|\Psi _{F}(\{u_{0}^{-+}\})\rangle .
\end{equation}

\textbf{(iii) PBC in }$n_{1}$\textbf{-direction and APBC in }$n_{2}$\textbf{-direction (PA-type):}

This case is similar to case (ii), so we obtain
\begin{equation}
Q_{c}|\Psi _{F}(\{u_{0}^{+-}\})\rangle =|\Psi _{F}(\{u_{0}^{+-}\})\rangle .
\end{equation}

\textbf{(iv) PBC in both directions (PP-type):}

For this case, the allowed lattice momenta satisfy $e^{i\mathbf{k}\cdot L_{1}\mathbf{n}_{1}}=e^{i\mathbf{k}\cdot L_{2}\mathbf{n}_{2}}=1$ with
\begin{eqnarray}
\mathbf{k}\cdot \mathbf{n}_{1} &=&0,\pm \frac{2\pi }{L_{1}},\pm \frac{4\pi }{L_{1}},\ldots ,\pm \frac{(L_{1}-2)\pi }{L_{1}},\pi ,  \notag \\
\mathbf{k}\cdot \mathbf{n}_{2} &=&0,\pm \frac{2\pi }{L_{2}},\pm \frac{4\pi }{L_{2}},\ldots ,\pm \frac{(L_{2}-2)\pi }{L_{2}},\pi ,
\end{eqnarray}
where all four TR-invariant points ($\mathbf{k}\cdot \mathbf{n}_{1}=0$,$\pi$ and $\mathbf{k}\cdot \mathbf{n}_{2}=0$, $\pi$) are allowed. According to Eq.~(\ref{eq:Qc-parity}), we should analyze their fermion parities separately. For ease of notation, we denote these points by $\mathbf{k}_{1}=(0,0)$, $\mathbf{k}_{2}$ ($\mathbf{k}_{2}\cdot \mathbf{n}_{1}=\mathbf{k}_{2}\cdot \mathbf{n}_{2}=\pi $), $\mathbf{k}_{3}$ ($\mathbf{k}_{3}\cdot \mathbf{n}_{1}=0$ and $\mathbf{k}_{3}\cdot \mathbf{n}_{2}=\pi $), and $\mathbf{k}_{4}$ ($\mathbf{k}_{4}\cdot \mathbf{n}_{1}=\pi $ and $\mathbf{k}_{4}\cdot \mathbf{n}_{2}=0$).

For the square lattice, the Hamiltonian $\tilde{H}+\tilde{H}^{\prime }$ [see Eqs.~(\ref{eq:H1}) and (\ref{eq:H2})] now has separate contributions from the TR-invariant points
\begin{eqnarray}
\tilde{H}+\tilde{H}^{\prime } &=&2i\sum_{\mathbf{k}}[e^{-i\mathbf{k}\cdot
\mathbf{n}_{1}}+e^{-i\mathbf{k}\cdot \mathbf{n}_{2}}+1-e^{-i\mathbf{k}\cdot (\mathbf{n}_{1}+\mathbf{n}_{2})}]c_{-\mathbf{k},\text{A}}c_{\mathbf{k},\text{B}}  \notag \\
&&-4i\kappa \sum_{\mathbf{k}}(e^{i\mathbf{k}\cdot \mathbf{n}_{1}}+e^{-i\mathbf{k}\cdot \mathbf{n}_{2}})c_{-\mathbf{k},\text{A}}c_{\mathbf{k},\text{A}}+4i\kappa \sum_{\mathbf{k}}(e^{i\mathbf{k}\cdot \mathbf{n}_{1}}+e^{-i\mathbf{k}\cdot \mathbf{n}_{2}})c_{-\mathbf{k},\text{B}}c_{\mathbf{k},\text{B}}  \notag \\
&=&\sum_{\mathbf{k}\in \mathrm{BZ}/2}c_{\mathbf{k}}^{\dagger }H(\mathbf{k})c_{\mathbf{k}}+4ic_{\mathbf{k}_{1},\text{A}}c_{\mathbf{k}_{1},\text{B}}-4ic_{\mathbf{k}_{2},\text{A}}c_{\mathbf{k}_{2},\text{B}}+4ic_{\mathbf{k}_{3},\text{A}}c_{\mathbf{k}_{3},\text{B}}+4ic_{\mathbf{k}_{4},\text{A}}c_{\mathbf{k}_{4},\text{B}}  \notag \\
&=&\sum_{\mathbf{k}\in \mathrm{BZ}/2}d_{\mathbf{k}}^{\dagger }\tilde{H}(\mathbf{k})d_{\mathbf{k}}+2i\tilde{c}_{\mathbf{k}_{1},\text{A}}\tilde{c}_{\mathbf{k}_{1},\text{B}}-2i\tilde{c}_{\mathbf{k}_{2},\text{A}}\tilde{c}_{\mathbf{k}_{2},\text{B}}+2i\tilde{c}_{\mathbf{k}_{3},\text{A}}\tilde{c}_{\mathbf{k}_{3},\text{B}}+2i\tilde{c}_{\mathbf{k}_{4},\text{A}}\tilde{c}_{\mathbf{k}_{4},\text{B}},
\end{eqnarray}%
where $\mathbf{k}\in \mathrm{BZ}/2$ contains $(N-4)/2$ points (with four TR-invariant points being excluded). This form of the Hamiltonian makes it convenient to calculate the fermion parities in its ground state $|\Psi_{F}(\{u_{0}^{++}\})\rangle $, where the $d_{\mathbf{k}}$-fermion parity is given by
\begin{equation}
(-1)^{\sum_{\mathbf{k}\in \mathrm{BZ}/2}(d_{\mathbf{k},1}^{\dagger }d_{\mathbf{k},1}+d_{\mathbf{k},2}^{\dagger}d_{\mathbf{k},2})}|\Psi_{F}(\{u_{0}^{++}\})\rangle =(-1)^{(N-4)/2}|\Psi _{F}(\{u_{0}^{++}\})\rangle
\end{equation}
due to the half filling and the fermion parity of the Majorana modes for the TR-invariant points are expressed as
\begin{eqnarray}
i\tilde{c}_{\mathbf{k}_{1},\text{A}}\tilde{c}_{\mathbf{k}_{1},\text{B}}|\Psi_{F}(\{u_{0}^{++}\})\rangle  &=&-|\Psi _{F}(\{u_{0}^{++}\})\rangle ,  \notag \\
i\tilde{c}_{\mathbf{k}_{2},\text{A}}\tilde{c}_{\mathbf{k}_{2},\text{B}}|\Psi_{F}(\{u_{0}^{++}\})\rangle  &=&|\Psi _{F}(\{u_{0}^{++}\})\rangle ,  \notag \\
i\tilde{c}_{\mathbf{k}_{3},\text{A}}\tilde{c}_{\mathbf{k}_{3},\text{B}}|\Psi_{F}(\{u_{0}^{++}\})\rangle  &=&-|\Psi _{F}(\{u_{0}^{++}\})\rangle ,  \notag \\
i\tilde{c}_{\mathbf{k}_{4},\text{A}}\tilde{c}_{\mathbf{k}_{4},\text{B}}|\Psi_{F}(\{u_{0}^{++}\})\rangle  &=&-|\Psi _{F}(\{u_{0}^{++}\})\rangle ,
\end{eqnarray}%
because of the energetic requirement. By using these results, we obtain the $c$-Majorana fermion parity for $|\Psi _{F}(\{u_{0}^{++}\})\rangle$
\begin{eqnarray}
Q_{c}|\Psi _{F}(\{u_{0}^{++}\})\rangle  &=&(-1)^{N/2}\exp \left( -i\frac{\pi}{2}\sum_{\mathbf{k}=\mathbf{k}_{1},\mathbf{k}_{2},\mathbf{k}_{3},\mathbf{k}_{4}}i\tilde{c}_{\mathbf{k},\text{A}}\tilde{c}_{\mathbf{k},\text{B}}\right)
(-1)^{\sum_{\mathbf{k}\in \mathrm{BZ}/2}(d_{\mathbf{k},1}^{\dagger}d_{\mathbf{k},1}+d_{\mathbf{k},2}^{\dagger }d_{\mathbf{k},2})}|\Psi_{F}(\{u_{0}^{++}\})\rangle   \notag \\
&=&(-1)^{N/2}\exp \left[ -i\frac{\pi }{2}(-1+1-1-1)\right] (-1)^{(N-4)/2}|\Psi _{F}(\{u_{0}^{++}\})\rangle   \notag \\
&=&-|\Psi _{F}(\{u_{0}^{++}\})\rangle ,
\end{eqnarray}%
which means that the $c$-Majorana fermion has an \textit{odd} fermion parity in $|\Psi _{F}(\{u_{0}^{++}\})\rangle $.

The analysis of the $c$-Majorana fermion for the honeycomb model is completely analogous and will not be repeated. One also obtains $Q_{c}|\Psi_{F}(\{u_{0}^{++}\})\rangle =-|\Psi _{F}(\{u_{0}^{++}\})\rangle$.

To summarize the results for all four cases, the fermion parity for a \emph{single} species of the itinerant Majorana fermions is \emph{even} for AA, AP, and PA boundary conditions, and \emph{odd} for the PP boundary condition. When taking into account all $\nu $\ copies of itinerant Majorana fermions, the fermionic ground state with the PP boundary condition would have fermion parity $(-1)^{\nu }$, while other three boundary conditions have even fermion parity
\begin{eqnarray}
Q_{1}|\Psi _{F}(\{u_{0}^{--}\})\rangle  &=&|\Psi_{F}(\{u_{0}^{--}\})\rangle,  \notag \\
Q_{1}|\Psi _{F}(\{u_{0}^{-+}\})\rangle  &=&|\Psi_{F}(\{u_{0}^{-+}\})\rangle,  \notag \\
Q_{1}|\Psi _{F}(\{u_{0}^{+-}\})\rangle  &=&|\Psi_{F}(\{u_{0}^{+-}\})\rangle,  \notag \\
Q_{1}|\Psi _{F}(\{u_{0}^{++}\})\rangle  &=&(-1)^{\nu }|\Psi_{F}(\{u_{0}^{++}\})\rangle .
\label{eq:itinerant-parity}
\end{eqnarray}

\subsubsection{$\mathbb{Z}_{2}$ gauge field}

Now we turn to the fermion parity for those Majorana fermions forming the static $\mathbb{Z}_{2}$ gauge field. For the ground-state subspace, these Majorana fermions can be viewed as stacked Kitaev's Majorana chains~\cite{kitaev2001} (with PBC or APBC depending on the four sectors) covering the rows and columns of the square and honeycomb lattices. When both $L_{1}$ and $L_{2}$ are even, it is easy to borrow the results from the Kitaev's Majorana chain to show that
\begin{equation}
Q_{2}|\{u_{0}^{\pm \pm }\}\rangle =|\{u_{0}^{\pm \pm }\}\rangle ,
\label{eq:gauge-parity}
\end{equation}
which holds for both square and honeycomb lattices.

By combining Eqs.~(\ref{eq:itinerant-parity}) and (\ref{eq:gauge-parity}), we obtain
\begin{eqnarray}
Q|\Psi _{F}(\{u_{0}^{--}\})\rangle \otimes |\{u_{0}^{--}\}\rangle  &=&|\Psi_{F}(\{u_{0}^{--}\})\rangle \otimes |\{u_{0}^{--}\}\rangle ,  \notag \\
Q|\Psi _{F}(\{u_{0}^{-+}\})\rangle \otimes |\{u_{0}^{-+}\}\rangle  &=&|\Psi_{F}(\{u_{0}^{-+}\})\rangle \otimes |\{u_{0}^{-+}\}\rangle ,  \notag \\
Q|\Psi _{F}(\{u_{0}^{+-}\})\rangle \otimes |\{u_{0}^{+-}\}\rangle  &=&|\Psi_{F}(\{u_{0}^{+-}\})\rangle \otimes |\{u_{0}^{+-}\}\rangle ,  \notag \\
Q|\Psi _{F}(\{u_{0}^{++}\})\rangle \otimes |\{u_{0}^{++}\}\rangle  &=&(-1)^{\nu }|\Psi _{F}(\{u_{0}^{++}\})\rangle \otimes
|\{u_{0}^{++}\}\rangle ,
\end{eqnarray}%
which proves that for \textit{odd} $\nu $, $|\Psi_{F}(\{u_{0}^{++}\})\rangle \otimes |\{u_{0}^{++}\}\rangle $ has an odd fermion parity and cannot survive the projection in Eq.~(\ref{eq:torus-degeneracy}).

\section{SO($\nu$) symmetry of the microscopic model}

In this section, we generalize the spin-orbital representation of the Hamiltonian, exemplified for $\nu=2$ and $\nu=3$ models in the main text, to arbitrary values of $\nu$. In this representation, the generalized spin sector has an explicit SO($\nu$) symmetry, with $\nu = 2q$ ($\nu = 2q+1$) for the model on the square (honeycomb) lattice and $q \in \mathbb{N}_0$.

We choose a representation in which the $2^{q+1}$-dimensional $\Gamma$ matrices satisfy
\begin{eqnarray}
\Gamma ^{\gamma 4} &=&\Lambda ^{1}\otimes \tau ^{\gamma },  \notag \\
\Gamma ^{\gamma 5} &=&\Lambda ^{2}\otimes \tau ^{\gamma },  \notag \\
&\vdots&   \notag \\
\Gamma ^{\gamma,2q+3} &=&\Lambda ^{2q}\otimes \tau ^{\gamma}, \notag \\
\Gamma ^{\gamma } &=&\Lambda ^{2q+1}\otimes \tau ^{\gamma },
\label{eq:gamma}
\end{eqnarray}
where $\gamma =1,2,3$, $(\tau ^{\gamma })_{\gamma =1,2,3}=(\tau^{x},\tau ^{y},\tau ^{z})$ are $2\times2$ Pauli matrices, and $\Gamma^{\alpha \beta} = \frac{i}{2} [\Gamma^\alpha, \Gamma^{\beta}]$ as in the main text.
The $\Lambda ^{a}$~($a=1,\ldots ,2q+1$) denote a $2^{q}$-dimensional representation of the Clifford algebra, satisfying $\{\Lambda ^{a},\Lambda^{b}\}=2\delta _{ab}$.
The representations of the matrices $\Gamma^4, \dots, \Gamma^{2q+3}$ can be obtained from Eq.~\eqref{eq:gamma} via
\begin{eqnarray}
\Gamma^4 & = & -i \Gamma^1 \Gamma^{14} = \Lambda^{1,2q+1} \otimes \mathbbm{1}, \notag \\
\Gamma^5 & = & -i \Gamma^1 \Gamma^{15} = \Lambda^{2,2q+1} \otimes \mathbbm{1}, \notag \\
& \vdots & \notag \\
\Gamma^{2q+3} & = & -i \Gamma^1 \Gamma^{1,2q+3} = \Lambda^{2q,2q+1} \otimes \mathbbm{1},
\end{eqnarray}
with $\Lambda^{ab} = \frac{i}{2} [\Lambda^a, \Lambda^b]$.
Following the $\nu =2$ and $\nu=3$ examples, we shall interpret $\Lambda ^{a}$ ($\tau ^{\gamma }$) as the generalized spin (orbital) degrees of freedom.
Note, however, that the matrices $\Lambda^a$ satisfy the spin algebra $[\Lambda^a, \Lambda^b] = i \epsilon^{abc} \Lambda^c$ only for $q=1$.

The $\nu=2q+1$ model on the honeycomb lattice can then be rewritten as
\begin{eqnarray}
H &=&-\sum_{\langle ij\rangle _{\gamma }}J_{\gamma }\left( \Gamma_{i}^{\gamma }\Gamma _{j}^{\gamma }+\sum_{\beta =4}^{2q+3}\Gamma_{i}^{\gamma \beta }\Gamma _{j}^{\gamma \beta }\right)
\notag \\
&=&-\sum_{\langle ij\rangle _{\gamma }}J_{\gamma }(\vec{\Lambda}_{i}\cdot\vec{\Lambda}_{j}) \otimes (\tau _{i}^{\gamma }\tau _{j}^{\gamma }),
\label{eq:spin-orbital}
\end{eqnarray}
where $\vec{\Lambda}\equiv (\Lambda ^{1},\Lambda ^{2},\ldots ,\Lambda ^{2q+1})$. Since $\vec{\Lambda}$ is a vector under SO($\nu$) with $\nu = 2q+1$, the Hamiltonian $H$ has an SO($\nu$) symmetry in the spin sector. It is also straightforward to show that the perturbation $H^{\prime }$ does not break the SO($\nu$) symmetry:
\begin{eqnarray}
H^{\prime } &=&-\kappa \!\sum_{\circlearrowright {\langle ijk\rangle }_{\gamma \gamma ^{\prime }}}\!\left( \Gamma _{i}^{\gamma }\Gamma
_{j}^{\gamma \gamma ^{\prime }}\Gamma _{k}^{\gamma ^{\prime }}-\sum_{\beta=4}^{2q+3} \Gamma _{i}^{\beta \gamma }\Gamma _{j}^{\gamma \gamma ^{\prime
}}\Gamma _{k}^{\gamma ^{\prime }\beta }\right)   \notag \\
&=&-i\kappa \!\sum_{\circlearrowright {\langle ijk\rangle }_{\gamma \gamma^{\prime }}}\!\left( \Gamma _{i}^{\gamma }\Gamma _{j}^{\gamma }\Gamma
_{j}^{\gamma ^{\prime }}\Gamma _{k}^{\gamma ^{\prime }}+\sum_{\beta=4}^{2q+3} \Gamma _{i}^{\gamma \beta }\Gamma _{j}^{\gamma }\Gamma_{j}^{\gamma ^{\prime }}\Gamma_{k}^{\gamma ^{\prime }\beta }\right)   \notag
\\
&=&-i\kappa \!\sum_{\circlearrowright {\langle ijk\rangle }_{\gamma \gamma^{\prime }}}\!(\Lambda _{i}^{2q+1}\Lambda _{k}^{2q+1}) \otimes (\tau _{i}^{\gamma }\tau_{j}^{\gamma } \tau _{j}^{\gamma ^{\prime }}\tau _{k}^{\gamma ^{\prime}})
-i\kappa \!\sum_{\circlearrowright {\langle ijk\rangle }_{\gamma \gamma
^{\prime }}}\sum_{a=1}^{2q} (\Lambda _{i}^{a}\Lambda _{k}^{a}) \otimes (\tau_{i}^{\gamma }\tau _{j}^{\gamma } \tau _{j}^{\gamma ^{\prime }}\tau_{k}^{\gamma ^{\prime }})  \notag \\
&=&\kappa \!\sum_{\circlearrowright {\langle ijk\rangle }_{\gamma \gamma^{\prime }}}\!(\vec{\Lambda}_{i}\cdot \vec{\Lambda}_{k}) \otimes (\tau _{i}^{\gamma} \tau _{k}^{\gamma^{\prime }}\tau _{j}^{\gamma''} ),
\end{eqnarray}
where $(\gamma, \gamma', \gamma'')$ is a permutation of $(1,2,3)$, such that $i$ and $j$ ($j$ and $k$) are connected via a link of type $\gamma$~($\gamma'$).

For the $\nu=2q$ model on the square lattice, one similarly arrives at
\begin{eqnarray}
H &=&-\sum_{\langle ij\rangle _{\gamma=1,\dots,4} } J_{\gamma }\left( \Gamma_{i}^{\gamma }\Gamma _{j}^{\gamma }+\sum_{\beta =5}^{2q+3}\Gamma_{i}^{\gamma \beta }\Gamma _{j}^{\gamma \beta }\right)
\notag \\
&=&-\sum_{\langle ij\rangle _{\gamma=1,2,3}}J_{\gamma }(\vec{\Lambda}_{i}\cdot\vec{\Lambda}_{j}) \otimes (\tau _{i}^{\gamma }\tau _{j}^{\gamma })
- \sum_{\langle ij\rangle _{\gamma=4}}J_{4}(\vec{\tilde \Lambda}_{i}\cdot\vec{\tilde \Lambda}_{j}) \otimes (\mathbbm{1} _{i} \mathbbm{1} _{j}),
\end{eqnarray}
where now $\vec{\Lambda}\equiv (\Lambda ^{2},\Lambda^{3},\ldots ,\Lambda ^{2q+1})$
and $\vec{\tilde\Lambda}\equiv (\Lambda^{12},\Lambda^{13},\ldots ,\Lambda ^{1,2q+1})$
which are both vectors under SO($\nu$) with $\nu = 2q$.
The Hamiltonian $H$ therefore has again an SO($\nu$) symmetry in the spin sector.

\bibliography{Supple}
\bibliographystyle{apsrev4-1}